\documentclass[10pt,conference]{IEEEtran}
\usepackage{framed,multicol}
\usepackage{xcolor,colortbl}
\usepackage{multirow}
\newcommand{\technique}{{\textsf BeAFix}}
\usepackage{listings} 
\definecolor{lightgray}{gray}{.60}
\definecolor{Lightgray}{gray}{.98}
\definecolor{bordegray}{gray}{.70}
\definecolor{gray}{gray}{.3}
\definecolor{commentcolor}{rgb}{0.33, 0.42, 0.18}

\lstdefinelanguage{alloylisting}{
  keywords={%
      assert, pred, all, no, lone, one, some, check, run,
      but, let, implies, not, iff, in, and, or, set, sig, Int, int,
      if, then, else, exactly, disj, fact, fun, module, abstract,
      extends, open, none, univ, iden, seq
  },
  sensitive=true,  
  morecomment=[l][\itshape \color{commentcolor}]//,
  morecomment=[l][\itshape \color{commentcolor}]{--},
  morecomment=[s][\itshape \color{commentcolor}] {/*}{*/},
  morestring=[b]",
  numbers=none,
  firstnumber=1,
  numberstyle=\tiny,
  stepnumber=1,
  numberstyle=\tiny\color{gray!50},
  numbersep=4pt,
  basicstyle=\footnotesize\ttfamily,
  commentstyle=\footnotesize\itshape,
  keywordstyle=\footnotesize\ttfamily\bfseries\color{black},
  ndkeywordstyle=\footnotesize\bfseries,
  frame=,
  framerule=.5pt,
  rulecolor=\color{bordegray!50},
  columns=fullflexible,
  keepspaces=true
}
\usepackage{verbatim}
\usepackage{flushend}

\begin{document}

\title{Bounded Exhaustive Search \\ of Alloy Specification Repairs}

\author{
\IEEEauthorblockN{Sim\'on Guti\'errez Brida\IEEEauthorrefmark{1}\IEEEauthorrefmark{2}, Germ\'an Regis\IEEEauthorrefmark{1}, Guolong Zheng\IEEEauthorrefmark{3}, \\ Hamid Bagheri\IEEEauthorrefmark{3}, ThanhVu Nguyen\IEEEauthorrefmark{3}, Nazareno Aguirre\IEEEauthorrefmark{1}\IEEEauthorrefmark{2}, Marcelo Frias\IEEEauthorrefmark{2}\IEEEauthorrefmark{4}}
\IEEEauthorblockA{\IEEEauthorrefmark{1}Department of Computer Science, FCEFQyN, University of R\'io Cuarto, Argentina} 
\IEEEauthorblockA{\IEEEauthorrefmark{2}National Council for Scientific and Technical Research (CONICET), Argentina} 
\IEEEauthorblockA{\IEEEauthorrefmark{3}Department of Computer Science \& Engineering, University of Nebraska-Lincoln, USA} 
\IEEEauthorblockA{\IEEEauthorrefmark{4}Department of Software Engineering, Buenos Aires Institute of Technology, Argentina}
}

\maketitle

\begin{abstract}

The rising popularity of declarative languages and the hard to debug nature thereof have motivated the need for applicable, automated repair techniques for such languages. However, despite significant advances in the program repair of imperative languages, there is a dearth of repair techniques for declarative languages. This paper presents {\technique}, an automated repair technique for faulty models written in Alloy, a declarative language based on first-order relational logic. {\technique} is backed with a novel strategy for bounded exhaustive, yet scalable, exploration of the spaces of fix candidates and a formally rigorous, sound pruning of such spaces. Moreover, different from the state-of-the-art in Alloy automated repair, that relies on the availability of unit tests, {\technique} does not require tests and can work with assertions that are naturally used in formal declarative languages. Our experience with using {\technique} to repair thousands of real-world faulty models, collected by other researchers, corroborates its ability to effectively generate correct repairs and outperform the state-of-the-art.

\end{abstract}


\section{Introduction}

Software has become ubiquitous, and many of our activities depend directly or indirectly on it. Having adequate software development techniques and methodologies that contribute to producing quality software systems has therefore become essential for many human activities. A well-established approach to achieving quality is to emphasize good problem understanding and planning ahead of development, i.e., to put an emphasis on the analysis and design phases of software development \cite{Ghezzi+2002}. These phases need to deal with descriptions of software and problem domains, which are typically captured using specification, or modeling, languages. Techniques and tools that allow users to \emph{analyze} specifications are very important, as they help developers in discovering flaws, such as missing cases in the specifications, wrong interpretations of requirements, etc. Two main problems arise in this phase: correctly \emph{understanding} the problem situation (thus capturing the right problem), and correctly \emph{stating} the problem in the language at hand (thus capturing the problem right). In the context of \emph{formal specification}, where formalisms with formal syntax and semantics are employed, the latter problem is particularly relevant, as the developer has to master the notation to correctly capture, in a formal way, a given software description \cite{ClarkeWing1996}. Even for experienced developers, many times subtle errors arise, like mistakenly using the wrong expression to capture a property, omitting an operator or using an operator in place of another, leading to incorrect specifications that do not capture the developer's intentions \cite{Nelson+2017}. These kinds of mistakes share characteristics with program defects. Therefore, techniques for dealing with these defects and, in general, to assess or improve software quality (such as techniques for bug finding and program debugging), are also relevant in the context of software specifications. In particular, techniques for improving debugging, e.g., via the automation of fault localization or program repair, are pertinent in the context of software specification. 

This paper targets the problem of automatically repairing formal specifications, more precisely, specifications in Alloy \cite{Jackson2006}, a formal language that has many applications in software development and has been successfully applied in a number of domains such as the discovery of design flaws in telecommunication applications \cite{Zave2017}, the analysis of security mechanisms in mobile and IoT platforms \cite{Bagheri+2018,DBLP:conf/issta/AlhanahnahSB20,DBLP:conf/dsn/BagheriSBM16}, the automation of software testing \cite{Khalek+2011,DBLP:conf/icse/MirzaeiGBSM16,Abad+2013}, and the verification of programs \cite{Dennis+2006,Galeotti+2010,Galeotti+2013}, among other applications \cite{Jackson2019}. While specifications share a number of characteristics with programs, certain characteristics make it non-trivial to apply the broad range of techniques for program repair, in the context of specifications. For instance, as a way to tame the space of candidates, various program repair techniques such as \emph{GenProg} \cite{LeGoues+2012} only use \emph{coarse-grained} syntactic modifications, such as block replacement, swapping, deletion and insertion, but no \emph{intra-statement} modifications are allowed. The rationale is that good levels of repairability in programs are achieved via coarse-grained modifications thanks to \emph{redundancies} that are present in code, especially in larger programs. Such redundancies are not often seen in specifications, in particular due to the relative conciseness of specifications compared to programs. Other approaches to program repair, e.g., \emph{PAR} \cite{Kim+2013}, restrict the modifications to patterns learned from human-written patches, mined from large repositories categorizing fixes; such inputs for the repair process are not available in the context of formal specification, simply because, as opposed to source code, there are no large repositories of specifications. Finally, most program repair techniques rely directly or indirectly on the availability of \emph{test cases}; while there exist initiatives that incorporate test cases to specifications \cite{Sullivan+2018}, other forms of checking, such as property satisfiability and verification, are more naturally found in specifications. 

In this paper, we present \technique, a novel technique that automatically repairs faulty Alloy specifications. \technique\ has several features distinguishing it from the state of the art \cite{Wang+2018}. Firstly, the technique does not depend on test cases, neither for fault localization nor for specification repair; it supports any kind of specification oracle, notably the typical assertion checks and property satisfiability checks found in Alloy specifications, as well as test cases. It is then more widely applicable in the context of formal specification, where test cases are rarely found accompanying specifications. Secondly, the technique tackles automated repair in a \emph{bounded exhaustive} way, i.e., by exhaustively exploring \emph{all} possible repair candidates, for a given set of mutation operators and maximum number of applications (on a set of identified suspicious specification locations). Thus, it either finds a fix, or guarantees that no fix is possible, within the provided bound and with the considered mutation operators over the identified faulty locations. This approach is natural to the context of Alloy, where users are accustomed to bounded exhaustive analyses.

\technique\ supports fine-grained mutations and is designed to enable the repair of multi-location specification defects. Since bounded exhaustive exploration suffers from inherent scalability issues, our technique features a number of \emph{pruning} strategies, that leverage the use of the Alloy Analyzer to \emph{soundly} prune large parts of the candidate space. More precisely, given a candidate repair for a specific suspicious location, our technique exploits both a syntactic analysis of the specification and a semantic analysis using the Alloy Analyzer for checking the feasibility of this candidate, in the sense that applying this specific repair candidate to the corresponding location preserves the feasibility of the overall (multi-location) repair. When feasibility fails, it allows us to prune, in a sound way, i.e., without losing valid fixes, significant parts of the search space for repair candidates, thus reducing specification repair running times. 

We evaluate our technique on a benchmark of Alloy specifications, including specifications previously used in assessing ARepair \cite{Wang+2018,Wang+2019}, and a large benchmark of faulty Alloy specifications produced by students \cite{Macedo+2020}. Our evaluation shows that our pruning technique significantly reduces specification repair running times, duplicating the number of repairs that can be produced within a 1-hour timeout, and reducing the repair time by 62X, on average. Moreover, when specifications feature typical assertions, and these are used as oracles, our technique shows a significant improvement in overfitting reduction, compared to the test-based technique ARepair.



\section{An Illustrating Example}

In this section, we introduce both Alloy and our technique by means of a motivating example. Alloy is a formal specification language, with a simple syntax and a relational semantics. The syntax of the language is rather small, and is compatible with an intuitive reading of specifications, or \emph{models}, as they are typically called in the context of Alloy \cite{Jackson2006} (we will use \emph{specification} and \emph{model} interchangeably in this paper). Specifications can resemble object-oriented notions that are familiar to developers. The basic syntactic elements of Alloy specifications are: \emph{signatures}, which declare data domains; \emph{signature fields} (akin to class attributes), that give structure to specifications and declare \emph{relations} between signatures; \emph{predicates}, parameterized formulas that can be used to state properties, represent operations, etc.; \emph{facts}, formulas that constrain the specifications and represent assumptions; and \emph{assertions}, formulas that capture \emph{intended} properties of the specification, i.e., properties that the user would like to verify. Formulas in Alloy are expressed in \emph{relational logic}, a first-order logic extended with relational operators such as relational transpose, union, difference and intersection. Alloy supports various quantifiers (\texttt{all} and \texttt{some} are the usual universal and existential quantifiers, respectively, \texttt{one} and \texttt{lone} are for ``exists exactly one'' and ``exists at most one'', respectively). It also features additional important relational operators: \emph{relational join}, a generalization of  composition to $n$-ary relations, which can be used to express \emph{navigations} as in object orientation; and \emph{transitive closure}, which can be applied only to binary relations, and extends the expressiveness of Alloy beyond that of first-order logic. 

\begin{figure}[ht!]
\begin{lstlisting} []
abstract sig Boolean { }
one sig True, False extends Boolean { }

sig Node {
 link: set Node,
 elem: set Int
}

sig List {
 header: set Node
}

fact CardinalityConstraints {
 all l : List | lone l.header
 all n : Node | lone n.link
 all n : Node | one n.elem
}

fact IGNORE {
 one List && List.header.*link = Node
}

pred Loop[This: List] {
 no This.header || 
 one n : This.header.*link | n.^link = n.*link 
}

pred Sorted[This: List] { // buggy
 all n: This.header.*link | n.elem < n.link.elem 
}

pred RepOk[This: List] {
 Loop[This] && Sorted[This]
}

run RepOk for 1 but exactly 3 Node expect 1

// buggy
pred Contains[This: List, x: Int, res: Boolean]{ 
 RepOk[This] &&
 ((x !in This.header.*link.elem => res=False ) || 
 res = True) 
}

pred Count[This: List, x: Int, res: Int] {
 RepOk[This] &&
 res = #{ n:This.header.*link | n.elem = x }
}

assert ContainsCorrect {
 all l : List, i, j : Int | 
   (Count[l, i, j] && j > 0) iff Contains[l, i, True]
}

check ContainsCorrect for 10
\end{lstlisting}
\caption{A (faulty) sample Alloy specification.}
\label{alloy-model}
\end{figure}

Consider the Alloy model in Figure~\ref{alloy-model}, a modified version of an Alloy specification of linked lists, that is part of the benchmark used in \cite{Wang+2018}. This model declares domains for booleans (with its two constants captured via singleton relations), and signatures for nodes and lists. Nodes have a link (a set of nodes), and associated elements (a set of integers); lists have a header (a set of nodes). A \emph{fact} constrains the cardinalities of these signature fields: lists have at most one header, and nodes have at most one successor node, and exactly one element (when applied to expressions, \texttt{lone}, \texttt{one} and \texttt{no} constrain a given expression to have a cardinality of at most one, exactly one, and exactly zero, respectively). Notice the additional fact, which is there for analysis purposes: it states that exactly one \texttt{List} is going to be considered in each instance of the model, and that all nodes present in an instance will be those in the list (no unreachable ``heap'' objects). Predicate \texttt{Loop} captures lists with a loop in its last node, saying that a list satisfies the predicate if it either has no header, or for exactly one of its nodes, the elements reachable in one or more steps from \texttt{link} are exactly the same reachable in zero or more steps through \texttt{link}. Predicate \texttt{Sorted} attempts to capture that lists are non-decreasingly sorted (this predicate is buggy though, as the order constraint is strict). Predicate \texttt{RepOk} is simply defined as the conjunction of \texttt{Loop} and \texttt{Sorted}. Predicate \texttt{Contains} is used to model an \emph{operation} on lists, namely, the operation for querying membership of an integer as an element of a node of a list. The result of the operation is captured by an additional Boolean parameter. This predicate is \emph{buggy}, it does not correctly model the intended operation (e.g., it admits the predicate to return \texttt{True} despite the contents of the list). 

Alloy specifications can be automatically analyzed, by an analysis mechanism that resorts to SAT solving, and is implemented in a tool called \emph{Alloy Analyzer} \cite{Jackson2006}. Two kinds of analysis are possible: \emph{running} a predicate and \emph{checking} an assertion. Both are analyzed in \emph{bounded} scenarios. Running a predicate searches for instances (scenarios) that satisfy all the constraints (cardinalities, facts, etc.), including the predicate being run. Assertion checking looks for \emph{counterexamples} of the asserted properties. Analysis is performed up to a bound $k$ (typically referred to as the \emph{scope} of the analysis), meaning, e.g., that assertion checking will either find a counterexample within the given scope, or guarantee the validity of the formula within the bound (similarly, a predicate will be found to be satisfiable within the provided scope, or not to have a satisfying instance within the scope). This \emph{bounded exhaustive analysis}, of course, does not necessarily mean that the formula is valid (resp., satisfiable), as counterexamples (resp., instances) of greater size may exist if larger scopes are considered.

The Alloy language is the vehicle for defining abstract software models in a lightweight and incremental way, with immediate feedback via automated analysis \cite{Jackson2006}. Typically, the process of constructing an Alloy model, as the one in our example, starts very much in the same way one would proceed while eliciting requirements, or sketching an abstract software design: basic domains of the model are identified (signatures of the model), over which more structured components are organized (signatures equipped with fields). How these domains and components are constituted, the inherent constraints of the problem domain and the operations that represent the software model capacities, are all incrementally created, via a constant interaction with the Alloy Analyzer. This process eventually involves the use of \emph{assertions} and \emph{predicates}, that capture intended properties of the model, and that serve essentially as the \emph{oracle} of the specification, i.e., the properties that would convey the acceptance of the model. Sometimes these properties can help find surprising counterexamples, that lead to refinements of the properties themselves, but more often they help one in ``debugging'' the core of the model, i.e., in getting the model ``right'', adapting it until the intended properties result as expected. For instance, for the linked lists model, the developer would expect the representation invariant \texttt{RepOk} to be satisfiable, and the definition of \texttt{Contains} to have the relationship with \texttt{Count} captured in property \texttt{ContainsCorrect}.  

While the intended properties are subject to defects too, they are typically significantly shorter and clearer than the ``core'' of the specification. They capture high level properties of the model, so they are expected to be simpler to write and get right. So, once the intended properties are set, the user may perform the corresponding analyses and use the results as an acceptance criterion for the specification, and the corresponding design it conveys. That is, a model will be considered incorrect if any of the analyses of the intended properties fails, i.e., has a result that contradicts the user expectations. In Figure~\ref{alloy-model}, for instance, the user may consider the consistency of \texttt{RepOk}, the assertion \texttt{ContainsCorrect} and the auxiliary predicate \texttt{Count} as the oracle of the specification, meaning that when this intended property is found to be invalid, the user would start modifying the remainder of the specification, as an attempt to fix the error. \technique\ as well as other model repair techniques aim at reducing human intervention along this overall modeling process, by automatically fixing errors in incorrect models.

Let us describe how the technique works, assuming for the moment that the faulty locations in the model have been correctly identified. In order to attempt to repair the specification, and assuming that for the first location the syntactic mutation operators lead to $n$ different fix candidates (for that specific location), and for the second location we have $m$ different fix candidates, in the worst case we have to check $n \times m$ potential fixes, as we would want to consider \emph{all} combinations of candidate fixes for each repair location. The model expectations, in our example the satisfiability of \texttt{RepOk} and the bounded validity of \texttt{ContainsCorrect}, will be the \emph{acceptance criterion} fix repair, i.e., if a fix candidate ``passes'' these analyses, it will be considered a fix.

The automated repair process for the above faulty specification is then straightforward to describe: we have $n \times m$ repair candidates (the combinations of fix candidates for the suspicious locations), and since we aim at exhaustively exploring this candidate space, we would run the oracles on each candidate, stopping as soon as we find one that ``passes'' all predicates and assertions. 

Let us describe some situations that allow for sound pruning, i.e., pruning that only avoids invalid fix candidates. 

Notice that, in our case, we have two defective lines, but these are not \emph{symmetric}: the bugs in \texttt{Sorted} affect \texttt{Contains}, as \texttt{Contains} depends on \texttt{RepOk} which in turn depends on \texttt{Sorted}, but the latter does not depend (i.e., calls directly or indirectly) on \texttt{Contains}. Thus, when checking a specific candidate for \texttt{Sorted} that does not pass an oracle involving \texttt{Sorted} but not \texttt{Contains}, as for instance the satisfiability of \texttt{RepOK}, we can stop analyzing the fix candidate for \texttt{Sorted} altogether, and not consider it in combination with any further candidates for the other location. Consider, for instance, the following combination of fix candidates for \texttt{Sorted} and \texttt{Contains}:

\begin{lstlisting} []
pred Sorted[This: List] {
    all n: This.header.*link | n.elem != n.link.elem
}

pred Contains[This: List, x: Int, res: Boolean] {
  RepOk[This] &&
  (x !in This.header.*link.elem => res = False) && 
  res = True
}
\end{lstlisting}

\noindent
Assuming that we consider the above described oracles for the specification, this combination does not pass the oracles, it is an invalid fix candidate. Moreover, if we leave the current fix candidate for \texttt{Sorted} and iterate over other candidates for \texttt{Contains}, the property check requiring \texttt{RepOk} to be satisfiable will continue to fail, as the unsatisfiability of \texttt{RepOk} cannot be solved by changing the definition of \texttt{Contains}. Thus, if we are able to identify this situation (as we explain later on, our technique does so), we can safely consider a different mutation for \texttt{Sorted}, or equivalently, soundly skip all combinations of the current mutation to \texttt{Sorted} with all other mutations for \texttt{Contains}. 

Now let us look at another situation, that will also allow us to soundly prune parts of the fix candidate space, even in the presence of bidirectional (or multi-directional) dependencies between faulty locations. Consider the above fix candidate for predicate \texttt{Contains}, that replaced \texttt{||} by \texttt{\&\&}. This ``local'' candidate that fails to pass an oracle such as the assertion on \texttt{Contains} (in combination with a particular candidate for \texttt{Sorted}) does not allow us to discard it altogether, as the failing cannot in principle be blamed on \texttt{\&\&} on its own: it may be the case that this candidate ``works'' with a different candidate for \texttt{Sorted}. So in order to check the local feasibility of the candidate for \texttt{Contains}, we need to consider it in combination with \emph{any} other candidate for \texttt{Sorted}, of course, trying to avoid checking \emph{all} candidates for this predicate. Assuming that we identified the body of the quantification of \texttt{Sorted} as the problematic part in that predicate (fault localization techniques for Alloy, in particular the one we use in this paper, can identify fine grained faulty locations, such as particular subexpressions), what we would need to intuitively check is whether there exists a (boolean) value for that location, that in combination with \texttt{\&\&} would make the oracles pass:

\begin{lstlisting} []
pred Sorted[This: List] {
    all n: This.header.*link | (??)
}

pred Contains[This: List, x: Int, res: Boolean] {
 RepOk[This] && 
  (x !in This.header.*link.elem => res = False) && 
  res = True
}

\end{lstlisting}

\noindent
That is, can we replace the double question mark above by a value that would make oracles pass? If the answer is \emph{no}, then we can blame \texttt{\&\&}, and try another candidate for \texttt{Contains}, avoiding considering of \texttt{\&\&} with candidates for \texttt{Sorted}. If we are able to correctly identify these situations, as our technique does and we describe later on in this paper, we can again safely prune a large number of candidates, namely all combinations of \texttt{\&\&} with all the mutations for \texttt{Sorted}.

It is worth remarking that we do not assume any particular format or characteristic, neither from the specification itself, nor from the oracle. This is in contrast with previous work on repairing Alloy specifications \cite{Wang+2018}, which requires repair oracles to be provided as Alloy \emph{test cases}. Alloy test cases define \emph{scenario-based expectations}, similar to what one would capture with unit tests for source code. As an example, consider the evaluation of \texttt{Contains} on a particular concrete structure, and its corresponding expected outcome (the expected outcome represents a boolean, 1 for ``satisfiable'' and 0 for ``unsatisfiable''): 

\begin{lstlisting} []
pred ContainsFalseOnListTest[This: List] {
  some n0, n1: Node | {
    This.header = n0 &&
    n0.link = n1 && n0.elem = 0 &&
    n1.link = n1 && n1.elem = 0 &&
    Contains[This, 1, False]
  }
}

run ContainsFalseOnListTest expect 1
\end{lstlisting}

\noindent
While scenarios do participate in the Alloy modeling process, they typically do so as a result of analyzing \emph{properties}. That is, tests are not a common explicitly described part of Alloy specifications. Recent proposals, notably \cite{Sullivan+2018}, are starting to motivate the use of test cases in formal specification. As mentioned, our approach allows for \emph{any} kind of oracle, including test-based oracles.

\section{The Technique}

Our approach to Alloy specification repair involves a series of tasks, for fault detection, fault localization, fix candidate generation, and fix candidate assessment. We describe these in more detail below. 

\subsection{Fault Detection and Fix Acceptance Criterion}

In general, given an Alloy specification, we may say that such specification is \emph{faulty} if at least one of the analysis commands in the specification has an outcome contrary to its corresponding expectation. This can be either a failing assertion (assertion with counterexamples), or a predicate that is unsatisfiable while the user expected it to be satisfiable, or vice versa. We may also allow for other flavors in commands, in particular Alloy test cases, in the spirit of AUnit \cite{Sullivan+2018}. The fault detection stage then resorts to SAT solving, the underlying analysis mechanism behind Alloy Analyzer, the tool for Alloy specification analysis \cite{Jackson2006}. Similarly, a fix candidate can be considered an acceptable patch when all the analysis commands in the specification have an outcome that coincides with the corresponding command's expectations. 

Our technique requires the user to identify the specification oracle, i.e., the assertions, predicates or tests that the technique will have to consider as fix acceptance criterion. The technique will then identify faults in the remainder of the specification (the oracle is left out of the analysis space for fault localization), and generate fix candidates for the faulty locations. Therefore, our repair approach cannot fix any faulty situation, but only those where the developer is certain about some part of it (the oracle), and wishes to alter the remainder of the specification to pass it. Looking for solutions that may modify the specification \emph{and} the criterion for acceptance would lead to fixes that may simply relax the acceptance criterion. Notice that, in this respect, we follow the same approach that ARepair and most test-based program repair techniques: the tests (the repair oracle) cannot be changed in the repair process. As described later on in this section, other trivial solutions such as changing a command's expectations or simply removing a command are prevented, due to how the fault localization is performed (which cannot be blamed on commands) and how fix candidates are generated (only by mutating the faulty locations). 

\subsection{Fault Localization}

Once a specification is deemed faulty, we need to identify the specific parts of the specification that are more likely to be blamed for the fault or faults. We do not deal with fault localization in this paper, and we assume an external technique/tool provides fault localization information. There exist techniques for fault localization that specifically target Alloy specifications, such as the spectrum-based fault localization mechanism behind ARepair \cite{Wang+2018}, and our fault localization technique presented in \cite{Zheng+2021}. While in principle any fault localization technique would fit our technique, as long as the employed fault localization can handle the oracles present in the faulty specification, it is worth to remark that the fault localization within ARepair inherently depends on having tests as oracles (acceptance criteria) for specifications \cite{Wang+2018}. Moreover, the fault localization in ARepair can dynamically change the identified faulty locations, as the specification is transformed during the repair process. Our technique, on the other hand, uses an \emph{offline} process for fault localization: the faulty program is fed to the fault localization tool, and a number of suspicious specification locations are returned. This is the input to our specification repair approach, and the space of \emph{all} possible patches for these locations, for a maximum depth in mutation application and a given set of mutation operators, will be considered. 

For our experiments in Section~IV, we use the FLACK fault localization technique \cite{Zheng+2021}. While we do not describe in detail the fault localization technique in this paper (we refer the reader to \cite{Zheng+2021}), let us remark a number of facts about FLACK: it supports arbitrary satisfiability checks and assertions, as well as tests, as specification oracles; it is based on the use of (partial) maximum satisfiability procedures, to process counterexamples of an Alloy model (witnessing the faulty status of the specification); and it can only identify faults within formulas and relational expressions, it cannot locate faults in data definitions, such as signature and field declarations, nor in commands (Alloy's runs and checks). 

\subsection{Generation of Fix Candidates}

Once the suspicious expressions are identified, syntactical variants of these expressions are produced. We consider an ample set of mutation operations, including the obvious logical and relational operator insertion, removal and replacement, quantification mutation (e.g., changing a quantifier), multiplicity constraint replacement, field/variable swap/replacement, etc., based on Alloy's grammar. Our tool processes the specification to obtain some typing information, so that some legal expressions that necessarily lead to empty relation/contradictory formulas are disregarded, as well as innocuous operation application (e.g., double transitive closure). Two elements are important to highlight here, namely the use of join to produce navigation chains, using fields, signatures, etc., and the possibility of \emph{combining} mutations, i.e., applying further mutations to an already mutated expression, akin the so called higher-order mutants \cite{DBLP:journals/infsof/JiaH09} in mutation testing.

Both the mutation operators and the maximum depth, i.e., the number of cumulative mutations (hence, the higher order nature of the generated mutants) that can be applied to a given faulty location, are configurable. These are bounded-exhaustively generated as the space of fix candidates is traversed (see below). In our experiments, we used 21 mutation operators in total, typically leading to roughly between 60 and 260 1-level mutants per location.  

\subsection{Fix Candidate Space Traversal}

Here we present our general repair approach. The two pruning techniques just introduced, are also described in more detail, and we argue about their soundness. The search space is organized as a search tree in a traditional search problem: the root is the original specification, with its faulty locations identified; and if a specification $s$ is in the tree and $s'$ can be obtained by applying a mutation to a faulty location, then $s'$ is also in the tree, with the same locations marked as faulty (so that the mutation process can be iterated). This in principle leads to an infinite fix candidate space, which we explore up to a maximum depth. While any search strategy may be applied, we explore the state space in a breadth-first fashion. 

\subsubsection{Partial repair checking}

Our first pruning strategy consists of identifying one of the suspicious locations for which a current repair candidate fails, as established by an analysis check that does not depend on the remainder of the faulty locations. We will describe it in more detail, assuming two faulty locations, without loss of generality. Let $\textit{Spec}$ be an Alloy specification, $\textit{Check}_1, \dots, \textit{Check}_k$ its analysis checks used as oracles, and $L_0, L_1$ the suspicious locations identified by the fault localization phase. Each analysis check $\textit{Check}_i$ refers to a specific part of $\textit{Spec}$, which can be determined by a straightforward syntactic analysis: $\textit{Check}_i$ refers to the formula it directly mentions (the body of the corresponding predicate or assertion), all the facts (axioms of the specification that are implicitly involved in every analysis check), and the symbols directly and indirectly referred syntactically to by these (predicates called, relations used, etc.). This syntactic analysis can determine then, for every $\textit{Check}_i$, which of the suspicious locations $L_0$ and $L_1$ it involves.

Most logics, and certainly Alloy's relational logic, have a sort of syntactic locality property, that guarantees that the validity/satisfiability of a formula depends only on the symbols it refers to. (In the case of Alloy, since validity/satisfiability is actually bounded validity/satisfiability, it can also depend on the scope, the bound, of analysis; but since the bound of analysis cannot be modified in the patch generation phase, we can disregard it). Moreover, the logic is monotonic, meaning that adding more assumptions to a formula can never reduce the conclusions drawn originally from it. These properties allow us to make the following observation. Let $m_0$ and $n_0$ constitute the modifications to locations $L_0$ and $L_1$, respectively, in the current fix candidate (i.e., be the expressions substituting the original expressions in locations $L_0$ and $L_1$ of $\textit{Spec}$). If a failing satisfiability check $\textit{Check}_i$ refers to only one of the suspicious locations, say $L_0$ and its current expression $m_0$, this means that the formula in $\textit{Check}_i$ is determined to be false independently of $n_0$. Then, for every alternative expression $n_i$ for location $L_1$, the corresponding fix candidate $(m_0, n_i)$ (the replacement expressions for locations $L_0$ and $L_1$) will still make $\textit{Check}_i$ to be false, due to the monotonicity of the logic. In other words, the specification cannot be repaired by modifying location $L_1$ if the current fix for location $L_0$ is maintained. We can therefore exclude (prune from the checking) all $(m_0, n_i)$ fix candidates as soon as we determine this situation, which in turn can be determined by a syntactic analysis of the specification, and the analysis outcome for fix candidate $(m_0, n_0)$. 

We refer to this analysis and the corresponding pruning it enables as \emph{partial repair checking}, due to the partiality of fix candidates when these do not involve all suspicious locations.

\subsubsection{Variabilization}

Our second pruning strategy is called \emph{variabilization}, due to the mechanism employed for prune checking, that requires introducing fresh variables to refer to fix candidates to specific locations, in a general way.

Let $\textit{Check}_i$ be a failing assertion (validity) check that refers to suspicious locations $L_0$ and $L_1$, and let $(m_0, n_0)$ be the current failing fix candidate. Notice that since $\textit{Check}_i$ is a failing validity check, we have a counterexample $\textit{CEX}_i$ as a result of the violation. That is, we have that:
\begin{displaymath}
    \textit{CEX}_i \not \models \textit{Spec}[m_0, n_0] \Rightarrow \textit{Check}_i,
\end{displaymath}
where $\textit{Spec}[m_0, n_0]$ denotes the fix candidate obtained by replacing $L_0$ and $L_1$ by $m_0$ and $n_0$, respectively, in $\textit{Spec}$. The purpose of variabilization is to check whether the current fix for $L_0$, i.e., $m_0$, \emph{may} work with \emph{some} candidate for $L_1$ (other than $n_0$, of course, which we already know it does not work). For technical reasons, we actually check whether some fix for $L_1$ may work in combination with $m_0$, for counterexample $\texttt{CEX}_i$. Let us describe the process for performing this check. 

Notice that fault locations can be subexpressions of a formula; let us refer by $F_1$ to the formula (predicate, fact, etc) containing $L_1$. Also, let $T$ be the most general type for $L_1$ in context $F_1$ (in Alloy, this most general type will depend on the arity required by $L_1$ in $F_1$, the context in which $L_1$ may depend upon, and will use the most general unary type, the universe \texttt{univ}). Let $\textit{Spec}_{L_1}$ be the specification obtained by replacing $F_1$ in $\textit{Spec}$ by 
\begin{displaymath}
\exists l_1: T \vert F_1[l_1/L_1] 
\end{displaymath}
i.e., we substitute $L_1$ by an existentially quantified variable of type $T$ (hence the name \emph{variabilization}). We now can check: 
\begin{displaymath}
    \textit{CEX}_i \models \textit{Spec}_{L_0}[m_0] \Rightarrow \textit{Check}_i,
\end{displaymath}
i.e., whether there exists a \emph{value} of type $T$ that can be put in place of location $L_1$, so that $\textit{CEX}_i$ ceases to be a counterexample. If this is the case, then local fix $m_0$ works as a fix for $L_0$, at least as far as $\textit{CEX}_i$ is concerned, and we may traverse the space of local candidates for $L_1$ to attempt to find a complete fix. But, on the other hand, if the above check fails, then there is no value that can be put in place of $L_1$ such that the local fix $m_0$ would work ($\textit{CEX}_i$ would still be a counterexample). Therefore, we can again exclude (prune from the checking) all $(m_0, n_i)$ fix candidates if the check fails. 

One may argue why not check the ``variabilized'' specification in the general case, instead of doing so \emph{only} for counterexample $\textit{CEX}_i$. The reason has to do with the type $T$ of location $L_1$. When this type is a relation of an arity greater than one, variabilization leads to a higher-order quantification, that Alloy cannot handle as a general analysis check, but it can do so for the specific counterexample $\textit{CEX}_i$. 

To clarify this variabilization process, and especially the reason why we typically have higher-order quantification, let us consider the example introduced in Section~II, where one local fix candidate is applied and the other was generalized with question marks. Assuming that assertion \texttt{ContainsCorrect} failed, a counterexample $\textit{CEX}$ was generated from this fix. To check whether variabilization pruning can be applied, we turn the question marks into existential quantifications. Intuitively, the corresponding variabilized specification would then be as follows (we are abusing the notation below, using Boolean for the type of the variabilized formula within \texttt{Sorted}):

\begin{lstlisting} []
pred Sorted[This: List] {
 some b: Boolean | all n: This.header.*link | b
}

pred Contains[This: List, x: Int, res: Boolean]{
 RepOk[This]
 (x !in This.header.*link.elem => res=False ) 
 && res = True
}
\end{lstlisting}

\noindent
However, we need to take into account that the variabilization context, the place where the location being variabilized occurs, depends in this case both on \texttt{This} and \texttt{n}. Thus, the actual variabilization for the check is as follows (we are again abusing the notation for the sake of clarity):

\begin{lstlisting} []
pred Sorted[This: List] {
 some b: List -> Node -> Boolean | 
     all n: This.header.*link | b[This, n]
}

pred Contains[This: List, x: Int, res: Boolean]{
 RepOk[This]
 (x !in This.header.*link.elem => res=False ) 
 && res = True
 \end{lstlisting}

\noindent
We cannot check \texttt{ContainsCorrect} over this specification due to the higher-order quantification in \texttt{Sorted}; but we can check it for $\textit{CEX}$. 

It is worth remarking that the above check, if successful, will produce a \emph{relational value} for \texttt{b} that makes the variabilized specification work. It will \emph{not} produce an expression to put in place of the body of the quantification, as a local fix candidate. It would not even produce a relational value that can be ``hardwired'' as a local fix of the corresponding location, since it is in principle just a relational value that works for counterexample $\textit{CEX}$. But its existence is what enables us to decide that a local fix for $L_1$ (\texttt{Sorted}) may be possible, considering the current local fix for $L_0$ (the \texttt{\&\&} in \texttt{Contains}). Our check essentially corresponds to only checking for feasibility of a local fix with respect to other locations. 

An alternative to the above would be to attempt to turn the relational value bound to $b$ into a relational \emph{expression}, that can be considered a local fix candidate. Such a process would correspond to a \emph{synthesis procedure}, which would require a grammar for expressions, so that the solver can attempt to work out an instance (an actual expression) during the satisfiability process. While it is technically feasible, it is also significantly more costly than our simpler query for satisfiability, which we solely use for pruning. 

\section{Evaluation}

We now assess our technique for automated repair of Alloy specifications. Our evaluation is based on two benchmarks of real faulty Alloy specifications, one taken from \cite{Nelson+2017} and used in the evaluation of ARepair \cite{Wang+2018}, and the other originated in the Alloy4Fun project \cite{Macedo+2020}, which includes 6 new models, with a total of 1936 faulty variants (considering different specification assignments resolved by different students). All the presented experiments were run on a 3.6GHz Intel Core i7 processor with 16 GB RAM, running GNU/Linux. We used a 1 hour timeout for each repair analysis instance. 

Our evaluation considers the following research questions:
\begin{itemize}

    \item \textbf{RQ1} What is the impact of the pruning strategies in the performance of our technique?

    \item \textbf{RQ2} How does our technique compare to previous work on automated repair for Alloy specifications?

\end{itemize}
For \textbf{RQ1}, notice that the pruning strategies only apply to specifications with multiple faulty locations. We then evaluate our technique, with pruning enabled vs. pruning disabled, over the following cases:
\begin{itemize}

    \item From ARepair's benchmark (we will refer in this way to the benchmark used in the original evaluation of \cite{Wang+2018}), we consider 18 specifications out of the 38 that are part of the benchmark. We disregard cases that have exactly one bug (20 in total in the benchmark), as these will not make pruning checks, nor trigger the pruning. 

    \item From Alloy4Fun, we consider a total of 273 faulty specifications. To build these specifications, we tracked the models with multiple assignments, and identified the cases in which a given model was submitted with more than one bug by the same student. (While the student id is not reported as part of the Alloy4Fun dataset, submissions are organized as chains of interaction ids, that correspond to a same student session. We use this information to organize submissions based on student sessions.) 

\end{itemize}
The results are summarized in Table~\ref{pruning-vs-no-pruning-experiments}. This table shows, for each of the benchmarks, the number of cases, how many were repaired with pruning enabled and disabled (recall the 1 hour timeout), and the average time for those cases that were repaired within the timeout (time is in milliseconds). We also report the increased repairability, and improved efficiency, obtained by pruning. We considered the cases that were not repaired with pruning disabled, but were repaired with pruning enabled, as if they were repaired in 1 hour. So, the increased efficiency is actually a lower bound of the actual improvement. For reference, we also report the range of efficiency improvement along all cases in each benchmark. 


\begin{table*}[ht]
\caption{Impact of pruning in repairability.}
\begin{center} 
\begin{small}
\begin{tabular}{|l|r|r|r|r|r|r|c|}
\hline
\rowcolor{black!25}
\cellcolor[gray]{.75}\textbf{Benchmark} & \cellcolor[gray]{.75}\textbf{Total} & \multicolumn{2}{c|}{\cellcolor[gray]{.75}\textbf{Pruning Disabled}} & \multicolumn{2}{c|}{ \cellcolor[gray]{.75}\textbf{Pruning Enabled}} & \multicolumn{2}{c|}{\cellcolor[gray]{.75}\textbf{Improved repairability/efficiency}}\\
\cellcolor[gray]{.75}{} & \cellcolor[gray]{.75}{\textbf{cases}} & \cellcolor[gray]{.75}{Repaired Cases} &\cellcolor[gray]{.75}\textit{Avg. Time} &\cellcolor[gray]{.75}{Repaired Cases} &\cellcolor[gray]{.75}\textit{Avg. Time} &\cellcolor[gray]{.75}{Repaired Cases} & \cellcolor[gray]{.75}\textit{Avg. Time \scriptsize{[Range min - max]}} \\
\hline
\multicolumn{8}{|c|}{\cellcolor[gray]{.90}\textit{\scriptsize{ARepair's benchmarks}}} \\
balancedBST & 2 & 0 & \textit{} & 0 & \textit{} & - & - \\ \hline
cd & 1 & 1 & \textit{1765} & 1 & \textit{540} & 1.00\scriptsize{\emph{X}} & 3\scriptsize{\emph{X}} [3\scriptsize{\emph{X}} - 3\scriptsize{\emph{X}}] \\ \hline
dll & 3 & 2 & \textit{290366} & 2 & \textit{2756} & 1.00\scriptsize{\emph{X}} & 80\scriptsize{\emph{X}} [26\scriptsize{\emph{X}} - 133\scriptsize{\emph{X}}] \\ \hline
farmer & 1 & 0 & \textit{} & 0 & \textit{} & - & - \\ \hline
fsm & 1 & 0 & \textit{} & 0 & \textit{} & - & - \\ \hline
student & 10 & 0 & \textit{} & 5 & \textit{184030} & 5.00\scriptsize{\emph{X}} & 26\scriptsize{\emph{X}} [1\scriptsize{\emph{X}} - 81\scriptsize{\emph{X}}] \\ \hline
\rowcolor{black!5}\textbf{Total:}  & 18 & 3 & \textit{146066} & 8 & \textit{62442} & 2.66\scriptsize{\emph{X}} & 37\scriptsize{\emph{X}} [1\scriptsize{\emph{X}} - 133\scriptsize{\emph{X}}] \\ \hline
\multicolumn{8}{|c|}{\cellcolor[gray]{.90}\textit{\scriptsize{Alloy4Fun's benchmarks}}} \\
Graphs & 22 & 6 & \textit{409667} & 16 & \textit{6821} & 2.66\scriptsize{\emph{X}} & 123\scriptsize{\emph{X}} [9\scriptsize{\emph{X}} - 387\scriptsize{\emph{X}}] \\ \hline
LTS & 33 & 0 & \textit{} & 1 & \textit{1983} & 1.00\scriptsize{\emph{X}} & 181\scriptsize{\emph{X}} [181\scriptsize{\emph{X}} - 181\scriptsize{\emph{X}}] \\ \hline
Trash & 23 & 7 & \textit{94960} & 15 & \textit{8084} & 2.14\scriptsize{\emph{X}} & 46\scriptsize{\emph{X}} [2\scriptsize{\emph{X}} - 107\scriptsize{\emph{X}}] \\ \hline
Production & 2 & 0 & \textit{} & 0 & \textit{} & - & - \\ \hline
Classroom & 169 & 14 & \textit{755978} & 32 & \textit{138447} & 2.28\scriptsize{\emph{X}} & 82\scriptsize{\emph{X}} [1\scriptsize{\emph{X}} - 433\scriptsize{\emph{X}}] \\ \hline
CV & 24 & 0 & \textit{} & 0 & \textit{} & - & - \\ \hline
\rowcolor{black!5}\textbf{Total:}  & 273 & 27 & \textit{420201} & 64 & \textit{38833} & 2.36\scriptsize{\emph{X}} & 85\scriptsize{\emph{X}} [1\scriptsize{\emph{X}} - 433\scriptsize{\emph{X}}] \\ \hline
\end{tabular}
\end{small}
\end{center}
\label{pruning-vs-no-pruning-experiments}
\end{table*}

For \textbf{RQ2}, we compare our technique with the only other approach for repairing Alloy models, namely ARepair \cite{Wang+2018}. We analyze both tools in their corresponding abilities to repair specifications in our considered benchmarks. For ARepair's benchmark, we used the models' corresponding assertions as oracles for \technique, and automatically generated test suites, using AUnit \cite{Sullivan+2018}, for ARepair. Recall that ARepair requires tests as oracles for the repair process; we actually follow the procedure suggested in \cite{Wang+2018}, as test cases are not commonly found accompanying Alloy specifications. Notice then that the results reported in \cite{Wang+2018} do not coincide with those reported here for ARepair's benchmark, as we use the same models with different test suites. The test suites used in \cite{Wang+2018} include manually designed cases, to help ARepair in overcoming overfitting. In our evaluation, we favored a comparison in which only the original assertions are available, and thus we generated test cases automatically, with AUnit (using the best performing criterion, predicate coverage \cite{Sullivan+2018}). 

From the Alloy4Fun dataset, we generated a benchmark consisting of: \emph{(i)} every faulty submission of the dataset as a single specification (these correspond to every intermediate specification submitted for analysis check in Alloy4Fun); and \emph{(ii)} the specifications combining all modifications within a single user session, that we used for \textbf{RQ1}. The total number of faulty specifications in this benchmark is 2209 (1936 faulty submissions, plus 273 sessions combining submissions of the same user). For \technique, we used the models' corresponding assertions as oracles. Since we do not have tests for these specifications, and ARepair inherently requires tests as repair oracles, we generated tests automatically using AUnit \cite{Sullivan+2018} (with predicate coverage as a target criterion), using the specification assertions, and employed these generated test suites for running ARepair. 

In all of the above cases, we contrasted the obtained repairs against correct versions of the corresponding specifications, using Alloy Analyzer, to account for overfitting. The results for ARepair and Alloy4Fun benchmarks are summarized in Tables~\ref{arepair-experiments} and \ref{alloy4fun-experiments}, respectively. For each model, we report the number of cases, and for each tool, the number of fixes found (percentage also reported), and how many of these are correct and incorrect (the latter, due to overfitting) patches. We also report the percentage of correct and incorrect patches, with respect to the total number of cases, and the average repair time in milliseconds, for each tool (these are the averages only for the repaired cases).




\begin{table*}[ht]
    \caption{Experiments taken from ARepair's benchmarks.} 
\begin{center} 
\begin{small}
\begin{tabular}{|l|r||r|r|r|r||r|r|r|r|}
\hline
\rowcolor{black!25}
\cellcolor[gray]{.75}\textbf{} & \cellcolor[gray]{.75}\textbf{Total} &\multicolumn{4}{|c||}{\textbf{ARepair }} &  \multicolumn{4}{|c|}{\textbf{ \technique } } \\
    \cellcolor[gray]{.75}\textbf{Model}& \cellcolor[gray]{.75}\textbf{Cases }& \multirow{ 2}{*}{\emph{Repaired} \scriptsize{(\%)}} &{\emph{Avg.}} & \multirow{ 2}{*}{\emph{Correct} \scriptsize{(\%)}}& \multirow{ 2}{*}{\emph{Incorrect} \scriptsize{(\%)}}& \multirow{ 2}{*}{\emph{Repaired} \scriptsize{(\%)}} &{\emph{Avg.}} & \multirow{ 2}{*}{\emph{Correct} \scriptsize{(\%)}}& \multirow{ 2}{*}{\emph{Incorrect} \scriptsize{(\%)}}\\
 \cellcolor[gray]{.75}\textbf{}& \cellcolor[gray]{.75}\textbf{}& \emph{} & \emph{time} & \emph{}& \emph{}& \emph{} & \emph{time} & \emph{}& \emph{} \\\hline\hline
addr & 1 & 1 \emph{\scriptsize(100\%)} & \textit{9010} & 1 \emph{\scriptsize(100\%)} & 0 \emph{\scriptsize(0\%)} & 1 \emph{\scriptsize(100\%)} & \textit{351} & 1 \emph{\scriptsize(100\%)} & 0 \emph{\scriptsize(0\%)} \\ \hline
arr & 2 & 2 \emph{\scriptsize(100\%)} & \textit{7651} & 2 \emph{\scriptsize(100\%)} & 0 \emph{\scriptsize(0\%)} & 2 \emph{\scriptsize(100\%)} & \textit{2394} & 2 \emph{\scriptsize(100\%)} & 0 \emph{\scriptsize(0\%)} \\ \hline
balancedBST & 3 & 2 \emph{\scriptsize(67\%)} & \textit{120276} & 1 \emph{\scriptsize(33\%)} & 1 \emph{\scriptsize(33\%)} & 1 \emph{\scriptsize(33\%)} & \textit{358} & 1 \emph{\scriptsize(33\%)} & 0 \emph{\scriptsize(0\%)} \\ \hline
bempl & 1 & 0 \emph{\scriptsize(0\%)} & \textit{} & 0 \emph{\scriptsize(0\%)} & 0 \emph{\scriptsize(0\%)} & 0 \emph{\scriptsize(0\%)} & \textit{} & 0 \emph{\scriptsize(0\%)} & 0 \emph{\scriptsize(0\%)} \\ \hline
cd & 2 & 2 \emph{\scriptsize(100\%)} & \textit{3302} & 0 \emph{\scriptsize(0\%)} & 2 \emph{\scriptsize(100\%)} & 2 \emph{\scriptsize(100\%)} & \textit{742} & 2 \emph{\scriptsize(100\%)} & 0 \emph{\scriptsize(0\%)} \\ \hline
ctree & 1 & 1 \emph{\scriptsize(100\%)} & \textit{6774} & 1 \emph{\scriptsize(100\%)} & 0 \emph{\scriptsize(0\%)} & 0 \emph{\scriptsize(0\%)} & \textit{} & 0 \emph{\scriptsize(0\%)} & 0 \emph{\scriptsize(0\%)} \\ \hline
dll & 4 & 3 \emph{\scriptsize(75\%)} & \textit{22585} & 0 \emph{\scriptsize(0\%)} & 3 \emph{\scriptsize(75\%)} & 3 \emph{\scriptsize(75\%)} & \textit{2624} & 3 \emph{\scriptsize(75\%)} & 0 \emph{\scriptsize(0\%)} \\ \hline
farmer & 1 & 0 \emph{\scriptsize(0\%)} & \textit{} & 0 \emph{\scriptsize(0\%)} & 0 \emph{\scriptsize(0\%)} & 0 \emph{\scriptsize(0\%)} & \textit{} & 0 \emph{\scriptsize(0\%)} & 0 \emph{\scriptsize(0\%)} \\ \hline
fsm & 2 & 2 \emph{\scriptsize(100\%)} & \textit{6068} & 2 \emph{\scriptsize(100\%)} & 0 \emph{\scriptsize(0\%)} & 1 \emph{\scriptsize(50\%)} & \textit{313} & 1 \emph{\scriptsize(50\%)} & 0 \emph{\scriptsize(0\%)} \\ \hline
grade & 1 & 1 \emph{\scriptsize(100\%)} & \textit{124797} & 0 \emph{\scriptsize(0\%)} & 1 \emph{\scriptsize(100\%)} & 0 \emph{\scriptsize(0\%)} & \textit{} & 0 \emph{\scriptsize(0\%)} & 0 \emph{\scriptsize(0\%)} \\ \hline
other & 1 & 0 \emph{\scriptsize(0\%)} & \textit{} & 0 \emph{\scriptsize(0\%)} & 0 \emph{\scriptsize(0\%)} & 1 \emph{\scriptsize(100\%)} & \textit{3120} & 1 \emph{\scriptsize(100\%)} & 0 \emph{\scriptsize(0\%)} \\ \hline
student & 19 & 12 \emph{\scriptsize(63\%)} & \textit{76120} & 9 \emph{\scriptsize(47\%)} & 3 \emph{\scriptsize(16\%)} & 13 \emph{\scriptsize(68\%)} & \textit{71197} & 13 \emph{\scriptsize(68\%)} & 0 \emph{\scriptsize(0\%)} \\ \hline
\rowcolor{black!5}\textbf{Total:}    & 38 & 26 \emph{\scriptsize(68\%)} & \textit{41843} & 16 \emph{\scriptsize(42\%)} & 10 \emph{\scriptsize(26\%)} & 24 \emph{\scriptsize(63\%)} & \textit{10137} & 24 \emph{\scriptsize(63\%)} & 0 \emph{\scriptsize(0\%)} \\ \hline
\end{tabular}
\end{small}
\end{center}
\label{arepair-experiments}
\end{table*}

\begin{table*}[ht]
    \caption{Experiments taken from Alloy4Fun's benchmarks.} 
\begin{center} 
\begin{small}
\begin{tabular}{|l|r||r|r|r|r||r|r|r|r|}
\hline
\rowcolor{black!25}
\cellcolor[gray]{.75}\textbf{} & \cellcolor[gray]{.75}\textbf{Total} &\multicolumn{4}{|c||}{\textbf{ARepair }} &  \multicolumn{4}{|c|}{\textbf{ \technique } } \\
    \cellcolor[gray]{.75}\textbf{Model}& \cellcolor[gray]{.75}\textbf{Cases }& \multirow{ 2}{*}{\emph{Repaired} \scriptsize{(\%)}} &{\emph{Avg.}} & \multirow{ 2}{*}{\emph{Correct} \scriptsize{(\%)}}& \multirow{ 2}{*}{\emph{Incorrect} \scriptsize{(\%)}}& \multirow{ 2}{*}{\emph{Repaired} \scriptsize{(\%)}} &{\emph{Avg.}} & \multirow{ 2}{*}{\emph{Correct} \scriptsize{(\%)}}& \multirow{ 2}{*}{\emph{Incorrect} \scriptsize{(\%)}}\\
 \cellcolor[gray]{.75}\textbf{}& \cellcolor[gray]{.75}\textbf{}& \emph{} & \emph{time} & \emph{}& \emph{}& \emph{} & \emph{time} & \emph{}& \emph{} \\\hline\hline
Graphs & 305 & 276 \emph{\scriptsize(90\%)} & \textit{2625} & 18 \emph{\scriptsize(6\%)} & 258 \emph{\scriptsize(85\%)} & 248 \emph{\scriptsize(81\%)} & \textit{6734} & 248 \emph{\scriptsize(81\%)} & 0 \emph{\scriptsize(0\%)} \\ \hline
LTS & 282 & 165 \emph{\scriptsize(59\%)} & \textit{8729} & 7 \emph{\scriptsize(2\%)} & 158 \emph{\scriptsize(56\%)} & 42 \emph{\scriptsize(15\%)} & \textit{5999} & 42 \emph{\scriptsize(15\%)} & 0 \emph{\scriptsize(0\%)} \\ \hline
Trash & 229 & 220 \emph{\scriptsize(96\%)} & \textit{4077} & 68 \emph{\scriptsize(30\%)} & 152 \emph{\scriptsize(66\%)} & 199 \emph{\scriptsize(87\%)} & \textit{4915} & 199 \emph{\scriptsize(87\%)} & 0 \emph{\scriptsize(0\%)} \\ \hline
Production & 63 & 47 \emph{\scriptsize(75\%)} & \textit{6232} & 8 \emph{\scriptsize(13\%)} & 39 \emph{\scriptsize(62\%)} & 56 \emph{\scriptsize(89\%)} & \textit{4124} & 56 \emph{\scriptsize(89\%)} & 0 \emph{\scriptsize(0\%)} \\ \hline
Classroom & 1168 & 911 \emph{\scriptsize(78\%)} & \textit{95717} & 92 \emph{\scriptsize(8\%)} & 819 \emph{\scriptsize(70\%)} & 418 \emph{\scriptsize(36\%)} & \textit{82856} & 418 \emph{\scriptsize(36\%)} & 0 \emph{\scriptsize(0\%)} \\ \hline
CV & 162 & 132 \emph{\scriptsize(81\%)} & \textit{4966} & 4 \emph{\scriptsize(2\%)} & 128 \emph{\scriptsize(79\%)} & 82 \emph{\scriptsize(51\%)} & \textit{2805} & 82 \emph{\scriptsize(51\%)} & 0 \emph{\scriptsize(0\%)} \\ \hline
\rowcolor{black!5}\textbf{Total:}      & 2209 & 1751 \emph{\scriptsize(79\%)} & \textit{20391} & 197 \emph{\scriptsize(9\%)} & 1554 \emph{\scriptsize(70\%)} & 1045 \emph{\scriptsize(47\%)} & \textit{17905} & 1045 \emph{\scriptsize(47\%)} & 0 \emph{\scriptsize(0\%)} \\ \hline
\end{tabular}
\end{small}
\end{center}
\label{alloy4fun-experiments}
\end{table*}

\subsection{Discussion}

Let us discuss the evaluation results. For \textbf{RQ1}, the results are conclusive: the impact of pruning is significant. Let us remark that the efficiency speed up is better than the increase in repairability (38X to 85X speed up, as opposed to roughly 2.5X increase in repairability). This may be explained by the timeout that we have set: 1 hour may be a small timeout for specification repair using \technique: increasing it may show a repairability increase closer to the speed up. Another important issue about these results is that the semantic check that we need to perform for pruning using variabilization, does in fact pay off. In other words, the variabilization checks, that require additional calls to the SAT solver, implied a time saving thanks to pruning that improved the overall analysis time. This, of course, is relative to the considered case studies. We did not observe any case where the overhead caused by pruning made the tool to actually take longer to repair a faulty specification, which may in fact happen for a specification, if most feasibility checks succeed, consuming time and leading to no pruning. The benchmarks were taken from other authors' work; we did not purposely look for specifications that may favor or harm the pruning strategies. We plan to design synthetic specifications, and extend the set of case studies, to further assess the effect of pruning.

Regarding \textbf{RQ2}, the comparison between \technique\ and ARepair can be analyzed along various dimensions. Let us first consider the evaluation over ARepair's benchmark. For this benchmark, the test suites used for running ARepair are solely composed of automatically generated tests, using AUnit with predicate coverage. As a result, the number of correct specification fixes differ from the experiments in \cite{Wang+2018}, where manually designed test cases helped the tool from overfitting. In our current experiments, ARepair is affected by overfitting: 16 out of the 26 produced fixes are correct fixes. \technique\ outperforms ARepair in terms of the number of repaired models: 16 models repaired by ARepair, against 24 repaired by \technique\ (a 21\% difference in the number of repaired models, over the size of the benchmark). It is worth remarking that the two techniques complement each other in terms of the repaired models: ARepair is able to repair models that \technique\ does not repair (see for instance \texttt{ctree} and \texttt{fsm}), and \technique\ repairs models that ARepair is not able to repair (see for instance \texttt{student} and \texttt{other}). In terms of efficiency, both tools show comparable running times. The average time to produce a repair is however just a reference, since the tools perform different kinds of tasks. \technique\ does not include fault localization, so the times here account for absolute repair times, given that the faults have been localized offline. ARepair, on the other hand, includes both the time to localize faults and perform the repair. Let us remark however that, in ARepair, on average 62\% of the time corresponds to repair and 38\% to fault localization. Unlike ARepair, that alternates between patching and calling fault localization, \technique\ calls fault localization only once, before triggering repair. As such, the proportion of time devoted to fault localization is much less. In our experiments, when we consider the combination of fault localization and {\technique}, on average 4\% is devoted to fault localization (in the worst case, Student6, the fault localization time was 13\% of the total time). Further details can be found in the tool's site (see below).  

Now let us consider the Alloy4Fun benchmark. For this benchmark, we did not have any choice but to automatically generate test cases, as these were not available for these models. We generated test cases automatically, using AUnit \cite{Sullivan+2018} (again, using the best performing generation criterion, as reported in \cite{Sullivan+2018}). ARepair is able to repair a significant number of models: 1751 out of 2209. However, only 197 were correct fixes; the remaining 1554 were overfitting cases, that passed the automatically generated tests, but were not correct fixes for the corresponding specifications. \technique, on the other hand, produced a smaller number of fixes: 1042 out of the 2209. But since it uses Alloy assertions as repair oracles, instead of test cases, it showed no overfitting issues for these specifications. As a result, \technique\ shows a better effectiveness in repair: 47\% of correctly repaired models by \technique, against 9\% of correctly repaired models by ARepair. Regarding the cases themselves, again, the tools complement each other: there are cases correctly repaired by one tool that were not repaired by the other, and vice versa. 

The observed overfitting is an important difference between the two tools and their approaches, and confirms our intuition and motivation regarding the use of stronger repair oracles, that naturally come in specifications. Clearly, one may argue that ARepair's performance, in terms of overfitting, can be improved by feeding the tool with different/stronger test suites. We fully agree, and in fact, this is confirmed with ARepair's benchmark: if the test suites used in \cite{Wang+2018} are fed to ARepair (which, as we mentioned, include manually crafted tests), then 26 out of 38 models are repaired, compared to the 14 out of 38 repaired models obtained with just automatically generated tests (effectiveness is increased from 42\% to 68\%). Writing the \emph{right} set of test cases for specification repair is a time consuming task, that would require a manual design of a test suite for each of the models, to improve the tool's results. The overfitting problem is an inherent problem of using tests as specifications, and thus it is expected of tools such as ARepair. 

It is important to remark that we do not claim that our technique leads to no overfitting, since this will depend on the oracle being used, and how faithfully it captures the developer's intentions. In the case of our controlled experiments, where we had the ground truths as oracles (which would not be the general case in formal specification), we had no overfitting, although overfitting may still have been observed due to the bounded nature of the analysis. In any case, being forced to use test cases as opposed to more general properties makes it more prone to overfitting. 

Other attributes of the generated patches may be considered. One of these is readability. We can remark that candidate patches are built out of mutations of the faulty expressions, and the space of faulty expressions is visited in breadth-first. Therefore, simpler/shorter fix candidates are considered first. While we did not evaluate readability in a systematic fashion, {\technique}'s patches can be simpler and clearer than manual, human-written ones. For instance, for Production.Inv4 in the Alloy4Fun benchmark, the faulty expression:
\begin{lstlisting} []
all c: Component | 
    (c.parts).position in (c.position).^~next
\end{lstlisting} 
is manually fixed by a student with the following expression: 
\begin{lstlisting} []
all c: Component | 
    ((c.^parts) & Component).position not in 
    (c.position).^next or no (c.^parts & Component)
\end{lstlisting}
\technique\, on the other hand, produces the following:
\begin{lstlisting} []
all c : one Component | 
    c.parts.position in c.position.~*next
\end{lstlisting}

Another dimension to consider is efficiency of our technique, compared with manual repairs. In Alloy4Fun we can measure the effort of human patches, by considering the time of the sessions of a same student, from defect introduction to its fixing. On average, it takes a student about 10 minutes to fix a defect, once it is introduced. On the other hand, the average time to repair in the case of \technique\ is about 10 seconds. For instance, for the above faulty specification, it took the student a total of 49 minutes to get it right. \technique\ repaired it in 3 seconds. Due to space reasons, we do not present here a more detailed comparison. The benchmarks, the tool's output with further statistical information, and the tool itself, can be found in the tool's site (see below).

\section{Related Work}

The problem of automatically repairing software defects has received great attention in the last decade, and a variety of techniques have been proposed to tackle it, including generate-and-validate techniques (e.g., based on evolutionary computation \cite{LeGoues+2012} or other forms of search in the space of candidates), techniques based on patch synthesis (e.g., techniques that gather constraints for correct program behavior and produce patches from these \cite{Mechtaev+2016}) and techniques driven by data (e.g., techniques based on learning \cite{LongRinard2016}). The emphasis is largely targeted at \emph{programs}, rather than \emph{specifications}. As explained earlier in this paper, the context of formal specification has some significant differences with programs (source code), that render many of these techniques not applicable, or at least difficult to adapt, to repairing specifications. The problem of dealing with the explosion of repair candidates has been dealt with in different ways, in the context of automated program repair. Some approaches attempt to bring down the branching factor in the search space by using a single mutation (e.g., \cite{DBLP:conf/tacas/GopinathMK11}); others consider a very small set of mutators (e.g., based on patterns of human-written fixes \cite{Kim+2013}), or consider coarse grained mutations (e.g., no intra-statement program modifications \cite{LeGoues+2012}). Most of these approaches perform non-exhaustive heuristic searches, as opposed to our technique, that proposes safely pruning the search space.

Our technique produces fine-grained repair candidates that are akin to \emph{mutations} \cite{AmmannOffutt2008}, such as operator and operand replacements, etc., or more generally, combinations of mutations (as in higher order mutations in the context of mutation testing \cite{JiaHarman2009}). The motivation for this decision is based on a number of issues, that seem to impact the effectiveness of larger-grained modifications (such as the copying, deletion and swapping of whole expressions) as operations to build repairs in the context of specification (for instance, for the case studies presented in \cite{Wang+2019}, our manual inspection showed no case where one may repair the specification by deleting, swapping or copying whole expressions within the specification). Firstly, specifications do not seem to feature the same level of reuse that programs have. For instance, in text books on formal specification with more traditional languages such as Z \cite{DBLP:books/daglib/0072139} or B \cite{Abrial2005}, one does not see modularization mechanisms (e.g., schema/machine composition) being used for reuse across different specifications, with the exception of the reuse of some general purpose specifications of sets, sequences, etc. Rather, modularization mechanisms seem to be exploited mainly for specification organization, with little impact in reuse. Secondly, most declarative specification languages are order-insensitive (the order of declarations and statements is irrelevant, as opposed to operational languages, making order-changing modifications ineffective). Thirdly, specifications are significantly shorter than source code, and therefore less redundancy that could be exploited for repairs is observed. 

While most work on automated repair applies to programs, there are some notable exceptions \cite{Pei+2014,Wang+2018,Wang+2019}. The tool AutoFix \cite{Pei+2014} targets contract-equipped programs, and can produce repairs that make the programs satisfy their contracts (at least as far as a test suite can determine). The technique can modify contracts as well as the code itself, and therefore can be considered as a specification repair technique. The approach differs from ours in many respects: it applies to specifications at the source-code level, as opposed to the more abstract specifications we target in this paper; it is not constrained to specifications, it can indistinguishably alter programs and specifications; and the specification is \emph{not} the oracle for repair, the tests are. An approach closely related to ours, as it applies to Alloy specifications too, is ARepair \cite{Wang+2018,Wang+2019}. ARepair repairs faulty Alloy specifications by combining a number of techniques, including a technique for synthesis known as \emph{sketching} \cite{DBLP:journals/sttt/Solar-Lezama13}, and mutation-based repairs, as in program repair. ARepair can fix specifications with multiple buggy locations, and is able to do so considering a manageable set of candidates, thanks to an effective fault localization approach (and resorting to sketching rather than arbitrary mutations). In effect, ARepair is guided by its own fault localization approach, and the whole process is supported by Alloy \emph{tests}. Our approach, on the other hand, is not coupled with fault localization, and can use different techniques (e.g., \cite{DBLP:journals/corr/abs-1807-08707,Zheng+2021}, as long as they can be used with the fault localization oracle at hand) for fault localization. Alloy tests are similar to unit tests for source code: they provide specific scenarios with an expected outcome when evaluating specific parts of an Alloy specification, e.g., a predicate. The tool has been successfully applied to repair specifications taken from a benchmark of Alloy models \cite{Nelson+2017} very efficiently, by being combined with techniques for automated Alloy test generation (as tests are necessary for repair). As for program repair techniques which use tests as acceptance criteria, they are subject to \emph{overfitting}, the problem that arises when a candidate passes all tests, but is not a true repair, i.e., there are situations in which the program (in this case, specification) fails to comply with the intended behavior. This, as usual, is strongly related to the quality of the provided test suite, and many of the cases from \cite{Nelson+2017} were repaired thanks to additionally, manually provided, test cases \cite{Wang+2018,Wang+2019}. ARepair inherently depends on test cases, while our technique works on arbitrary Alloy specification oracles. See the previous section for a more detailed comparison of \technique\ with ARepair, from a more experimental point of view. 

Our technique uses Alloy counterexamples to weakly check variabilization feasibility, since fully checking feasibility requires dealing with higher-order quantification. To perform this higher-order checking, one may use Alloy* \cite{DBLP:journals/fmsd/MilicevicNKJ19}. We experimented with this approach, but due to performance issues, we favored our current counterexample-based mechanism. Also in this line, one may profit from Alloy* to capture Alloy's grammar and semantics into Alloy*, and use the solver to encode the whole repair approach. In this way, Alloy* would function as a synthesis engine, with the solver doing the search for repairs, as in some semantic program repair approaches (e.g., \cite{Mechtaev+2016}). In our initial attempts we did not manage to obtain results, due to the available heap space being exceeded, for fragments of Alloy's grammar significantly smaller than what we are considering with our ad-hoc search approach. We plan however to further investigate this possibility.

\section{Conclusion}

Software specification and modeling are crucial activities of most software development methods. Getting a software specification right, i.e., capturing correctly a software design, the constraints and expected properties, etc., especially when the language to capture these is \emph{formal}, is very challenging. Thus, techniques and tools that help developers in correctly specifying software is highly relevant. In this paper, we have presented a technique that helps precisely in this task, in the context of formal specification using the Alloy language \cite{Jackson2006}. Our technique has a number of characteristics that distinguish it from related work \cite{Wang+2019}. Firstly, it does not require any particular form of the \emph{oracles}, i.e., the properties to be used for assessing fix candidates (as opposed to existing work which require such oracles to be expressed in terms of test cases). Secondly, it bounded exhaustively explores the state space of fix candidates, thus finding a specification fix, or guaranteeing that such a fix is impossible within the established bounds, for the identified faulty locations, and with the provided mutation (syntactic modification) operators. This is suitable in an Alloy context, where users are accustomed to bounded-exhaustive analyses. This bounded-exhaustive exploration of fix candidates demands then appropriate mechanisms to make the search more efficient. Our technique comes with two sound pruning strategies, that allow us to avoid visiting large parts of the state space for fix candidates, which are guaranteed not to contain valid fixes. We have assessed our technique on a large benchmark of Alloy specifications, and shown that the pruning strategies have an important impact in analysis. The technique has an efficiency comparable to that of the previous work  \cite{Wang+2019}, it complements the latter in terms of the fixes it is able to generate, and is less prone to overfitting, as it naturally supports stronger oracles based on assertion checking and property satisfiability, that usually accompany Alloy specifications.

\section{Data Availability}
{\technique}, all benchmark data, further statistical information and the instructions to replicate the experiments in this paper, are available at \cite{beafix-site}. A snapshot of the tool and benchmark, as used in the paper, is available at \cite{beafix-snapshot}.

\section*{Acknowledgments}
We thank the anonymous reviewers for their helpful comments. This work was supported in part by awards W911NF-19-1-0054 from the Army Research Office; CCF-1948536, CCF- 1755890, and CCF-1618132 from the National Science Foundation; and PICT 2016-1384, 2017-1979 and 2017-2622 from the Argentine National Agency of Scientific and Technological Promotion (ANPCyT).

\newpage

\bibliographystyle{IEEEtran}
\bibliography{bibliography}

\end{document}